\documentclass[doublecol]{epl2}
\usepackage{amsmath}
\usepackage{amssymb}
\usepackage{amsfonts}

\def\nl{\\ & \quad}

\allowdisplaybreaks

\begin{document}
\title{Canonical formulation of self-gravitating spinning-object systems}
\date{\today}
\author{Jan Steinhoff and Gerhard Sch\"afer}
\shortauthor{J.~Steinhoff and G.~Sch\"afer}
\institute{Theoretisch-Physikalisches Institut, Friedrich-Schiller-Universit\"at, Max-Wien-Platz\ 1, 07743 Jena, Germany, EU}

\abstract{
Based on the Arnowitt-Deser-Misner (ADM) canonical
formulation of general relativity, a canonical formulation of
gravitationally interacting classical spinning-object systems
is given to linear order in spin. The constructed position,
linear momentum and spin variables fulfill
standard Poisson bracket relations. A spatially symmetric time gauge
for the tetrad field is introduced. The achieved formulation is of
fully reduced form without unresolved constraints, supplementary,
gauge, or coordinate conditions. The canonical field momentum
is not related to the extrinsic curvature of spacelike hypersurfaces
in standard ADM form. A new reduction of the tetrad degrees of freedom
to the Einstein form of the metric field is suggested.
}
\maketitle

The canonical formulation of spinning objects under gravitational interaction
is an important issue in general relativity. For the gravitationally
interacting Dirac field, e.g., several
investigations have been undertaken to settle the problem,
\cite{D62,K63,DI76,GH77,NT78,H83}.
For spinning classical objects like rotating black holes, neutron stars,
or other stars similar advances have not yet been
obtained.  For objects in external gravitational fields, canonical formulations
were constructed by K\"unzle \cite{K72} and, very recently, by Barausse et al. \cite{BRB09}.
In the paper by Yee and Bander \cite{YB93} a Routhian approach was developed with
the spin part in canonical form. Within effective field theory
techniques, the latter approach was pushed forward by Porto and
Rothstein \cite{P08} to self-gravitating spinning objects operating in
non-reduced phase space of the matter variables with the reduction to be
performed within post-Newtonian (PN) approximations. The formalism was
shown to operate well to the next-to-leading order spin-spin
coupling. Based on the Arnowitt-Deser-Misner (ADM) canonical
formulation of general relativity, \cite{ADM62}, starting from the stress-energy
tensor for pole-dipole objects by Tulczyjew and Dixon, a fully reduced
matter-field canonical formalism was achieved, but to next-to-leading
order in the PN approximation only \cite{SSH08}. Most recently, this approach has been
generalized to a next-order PN approximation \cite{SW09}. Within
the previously developed canonical scheme even quadratic-in-spin
interactions could be treated successfully, \cite{SHS08}.
The leading order spin-gravity interaction has found various derivations, for quantum
and classical spinning particles see, e.g., \cite{BC79}, for black
holes and other bodies, \cite{DE75}. With the aid of a test-spin-type
approach, the next-to-leading order spin-orbit dynamics in canonical
form has been obtained for the first time only recently,
\cite{DJS08}. It has turned out fully equivalent to the non-canonical result
in \cite{FBB06}.

In the present article, a canonical formulation for self-gravitating spinning
classical objects will be developed which, to linear order in the
spin, is valid for arbitrary gravitational fields. The formulation
makes crucial use of a spatially symmetric time gauge for the tetrad field
similar to the one introduced by Kibble \cite{K63} and, like in Kibble's investigation of the
gravitationally interacting Dirac field, a full reduction is achieved
without remaining unresolved constraints. The gravitational
field turns out to be treated within a field-momentum-generalized version of the
ADM canonical formalism. Thus, about 50 years after the seminal work
by ADM, classical spinning-object systems receive a canonical
implementation into the Einstein theory of gravity.
The importance of a canonical formulation is discussed, e.g., in \cite{IN80}.
Most notably it allows a thorough analysis of the consistency
of the system of equations consisting of the Mathisson-Papapetrou equations, (\ref{EOM1}) and (\ref{EOM2}),
and the Einstein equations with the Tulczyjew-Dixon stress-energy tensor (\ref{SET})
as a source. (This system is applicable to the dynamics of compact
objects like black holes or neutron stars if rotation is not too rapid,
i.e., if spin-squared terms can be neglected, and if tidal deformation
has no effect. It was thus the basis for many investigations of such
objects in the past. Of course, also weakly gravitating objects can be
modeled in this way as long as their deformation is negligible.)
In particular, the action used in the present paper belongs to the class of
so-called derivative coupled theories, which are
known to potentially reveal subtle inconsistencies when formulated canonically, see \cite{IN80}.
However, for the Dirac field the problematic terms containing the extrinsic curvature of the space-like
hypersurfaces can be eliminated by a redefinition of the Dirac field.
In our case, those problematic terms are absorbed into the canonical momentum of the matter.
Besides rather mathematical consistency considerations that now become available,
applications of the canonical formulation given in the present paper to the
post-Minkowskian (see, e.g., \cite{LSB08}) and to higher order PN approximations
(next-to-next-to-leading order spin-obit in particular) are most interesting.
Further, the suggested gravitational-field reduction will apply
to the Dirac field as well with possibly nice prospects for
applications, cf., \cite{OST09}.

Latin indices from the middle of the alphabet
are running through $i = 1, 2, 3$. We utilize three different frames here,
denoted by different indices. Greek indices refer to the coordinate frame
and have the values $\mu = 0, i$.
Lower case Latin indices from the beginning of the alphabet refer to the
local Lorentz frame, while upper case ones denote the so called
body-fixed Lorentz frame. The values of these Lorentz indices are marked by round and
square brackets as $a = (0), (i)$ and $A = [0], [i]$, respectively, e.g., $A = [0], [1], [2], [3]$.
Partial derivatives are denoted $~_{,\mu}$.
Both the speed of light, $c$, and the Newton gravitational constant, $G$, are put equal to 1.

Our starting point will be an action functional $W$ which is
invariant against general four-dimensional coordinate transformations, general local
tetrad rotations, and reparametrization of the objects affine time parameter.
In terms of a Lagrangian density $\mathcal{L}$ the action reads
\begin{equation}
W[e_{a \mu}, z^{\mu}, p_{\mu}, \Lambda^{Ca}, S_{ab}, \lambda_1^a, \lambda_{2[i]}, \lambda_3]
	= \int \mathrm{d}^4 x \, \mathcal{L} \,,
\end{equation}
and must be varied with respect to the tetrad field $e_{a \mu}$,
the Lagrange multipliers $\lambda_1^a$, $\lambda_{2[i]}$, $\lambda_3$,
position $z^{\mu}$ and linear momentum $p_{\mu}$ of the object,
as well as with respect to angle-type variables $\Lambda^{Ca}$ and spin tensor $S_{ab}$
associated with the object. The angle variables are
represented by a Lorentz matrix satisfying
$\Lambda^{Aa} \Lambda^{Bb} \eta_{AB} = \eta^{ab}$ or
$\Lambda_{Aa} \Lambda_{Bb} \eta^{ab} = \eta_{AB}$, where $\eta_{AB} = \text{diag}(-1,1,1,1) = \eta^{ab}$,
which must be respected upon variation, see \cite{HR74}.
The matter part of the Lagrangian density reads
\begin{equation}
\mathcal{L}_M = \int \mathrm{d}\tau \left[ \left( p_{\mu}
- \frac{1}{2} S_{ab} \,\omega_{\mu}^{~ab}\right) \frac{\mathrm{d} z^{\mu}}{\mathrm{d}\tau}
+ \frac{1}{2} S_{ab} \frac{\mathrm{d} \theta^{ab}}{\mathrm{d}\tau} \right] \delta_{(4)} \label{spin1} \,,
\end{equation}
with $\mathrm{d} \theta^{ab} = \Lambda_{C}^{~a} \mathrm{d} \Lambda^{Cb} = - \mathrm{d} \theta^{ba}$ and
$\delta_{(4)} = \delta(x^{\nu}-z^{\nu}(\tau))$ the 4-dim.\ delta
function, $\int \mathrm{d}^4x\,\delta_{(4)}=1$. The objects affine time variable is $\tau$.
The Ricci rotation coefficients $\omega_{\mu}^{~ab}$ are given by
$\omega_{\mu\alpha\beta} = e_{a\alpha} e_{b\beta} \omega_{\mu}^{~ a b}
	= - \Gamma_{\beta\alpha\mu}^{(4)} + e_{\alpha , \mu}^c e_{c \beta}$,
with $\Gamma_{\beta\alpha\mu}^{(4)} = \frac{1}{2} ( g_{\beta\alpha,\mu} + g_{\beta\mu,\alpha} - g_{\alpha\mu,\beta})$
the 4-dim.\ Christoffel symbols of first kind and $g_{\mu\nu} = e_{a\mu} e_{b\nu} \eta^{ab}$
the 4-dim.\ metric. As in \cite{HR74}, the matrix $\Lambda^{Ca}$ can be subjected
to right (or left) Lorentz transformations, which correspond to
transformations of the reference frame (or the body-fixed frame).
The spin part of $\mathcal{L}_M$ can be obtained from the kinetic term
$\frac{1}{2} S_{ab} \frac{\mathrm{d} \theta^{ab}}{\mathrm{d}\tau}$ in Minkowski
space, see \cite{HR74}, by promoting the global symmetry of this term under right
Lorentz transformations (i.e., transformations of the reference frame),
to a local symmetry, a usual procedure for the construction of gauge theories.
This is achieved by introducing the $\omega_{\mu}^{~ab}$ term in (\ref{spin1}).
The matter constraints are given by
\begin{equation}
\mathcal{L}_C = \int \mathrm{d}\tau \left[
	\lambda_1^a p^b S_{ab} + \lambda_{2[i]} \Lambda^{[i] a} p_a
	 - \frac{\lambda_3}{2} ( p^2 + m^2 ) \right] \delta_{(4)} \,,
\end{equation}
where $m$ is the constant mass of the object and $p^2 = p_{\mu} p^{\mu}$.
The constraint $p^b S_{ab} = 0$ (spin supplementary condition, SSC)
states that in the rest frame the spin tensor contains the
3-dim.\ spin $S_{(i)(j)}$ only (i.e., the mass dipole part $S_{(0)(i)}$ vanishes),
while the conjugate constraint $\Lambda^{[i] a} p_a = 0$ ensures
that $\Lambda^{C a}$ is a pure 3-dim.\ rotation matrix in the rest frame
(no Lorentz boosts), see \cite{HR74}.
Finally, the gravitational part is given by the usual
second order Einstein-Hilbert Lagrangian density
\begin{equation}\label{EH}
\mathcal{L}_G = \frac{1}{16\pi} \sqrt{-g} \text{R}^{(4)} + (\text{TD}) \,,
\end{equation}
where $g$ is the determinant of the 4-dim.\ metric and $\text{R}^{(4)}$
is the 4-dim.\ Ricci scalar. Using a second order form of the
gravitational action, i.e., not varying the connection
independently, ensures that the torsion tensor must vanish, see, e.g., 
\cite{NT78}. A total divergence in the Lagrangian density may be added without affecting
the equations of motions, as they can be obtained from local variations.
The term denoted $(\text{TD})$ is written here as a reminder of this
fact, and will become important later on. The complete Lagrangian density is the sum
\begin{equation}
\mathcal{L} = \mathcal{L}_G + \mathcal{L}_M + \mathcal{L}_C \,.
\end{equation}
We assume asymptotical flatness as a boundary condition of the spacetime.

Variation of the action $\delta W = 0$ leads to the equations of motion (EOM) for the matter variables
\begin{gather}
	\frac{\mathrm{D} S_{ab}}{\mathrm{D} \tau} = 0 \,, \quad
	\frac{\mathrm{D} \Lambda^{Ca}}{\mathrm{D}\tau} = 0 \,, \quad
	u^{\mu} \equiv \frac{\mathrm{d} z^\mu}{\mathrm{d} \tau} = \lambda_3 p^{\mu} \,, \label{EOM1} \\
	\frac{\mathrm{D} p_{\mu}}{\mathrm{D} \tau} = - \frac{1}{2} \text{R}_{\mu\rho ab}^{(4)}
		u^{\rho} S^{ab} \,, \label{EOM2}
\end{gather}
as well as to the usual Einstein equations with
the stress-energy tensor
\begin{align}
T^{\mu\nu} &= \frac{e^{\mu}_a}{\sqrt{-g}} \frac{\delta ( \mathcal{L}_M + \mathcal{L}_C )}{\delta e_{a \nu}} \\
&= \int \mathrm{d} \tau \left[
	\lambda_3 p^{\mu} p^{\nu} \frac{\delta_{(4)}}{\sqrt{-g}}
	+ \bigg( u^{(\mu} S^{\nu)\alpha} \frac{\delta_{(4)}}{\sqrt{-g}} \bigg)_{||\alpha}
\right] \,, \label{SET}
\end{align}
where $\text{R}_{\mu\rho ab}^{(4)}$ is the 4-dim.\ Riemann tensor,
$_{||\alpha}$ denotes the 4-dim.\ covariant derivative,
and $\mathrm{d}$ and $\mathrm{D}$, respectively, denote ordinary and covariant total derivatives.
Here it was already used that preservation of the constraints
in time requires $\lambda_1^a$ to be proportional to $p^a$
and $\lambda_{2[i]}$ to be zero, so that $\lambda_1^a$ and $\lambda_{2[i]}$
drop out of the matter EOM and the stress-energy tensor.
The Lagrangian multiplier $\lambda_3 = \lambda_3(\tau)$ represents the reparametrization
invariance of the action (notice $\lambda_3 = \sqrt{-u^2}/m$).
Further, an antisymmetric part of the stress-enery tensor
\begin{equation}
\int \mathrm{d} \tau \left( \frac{1}{2} S^{\mu\nu} u^{\rho} \frac{\delta_{(4)}}{\sqrt{-g}} \right)_{||\rho} =
	\int \mathrm{d} \tau \frac{1}{2} \frac{\mathrm{D} S^{\mu\nu}}{\mathrm{D} \tau} \frac{\delta_{(4)}}{\sqrt{-g}} = 0
\end{equation}
vanishes and it holds $T^{\mu\nu}_{~~||\nu} = 0$ by virtue of the matter EOM. Obviously, the spin
length $s$ defined by $2 s^2 = S_{ab}S^{ab}$ is conserved.

To our knowledge, the Lagrangian density
in the form given here is used for the first time, though the expressions
applied in, e.g., \cite{K72,YB93,P06,P08} are somewhat
related. In \cite{K72,YB93} spinning objects in an external gravitational field
are treated, in contrast to self-gravitating spinning objects here.
In \cite{P06} the metric is the fundamental
variable (e.g., used in functional integrations)
instead of the tetrad field, and \cite{P08} gives a Routhian version of \cite{P06}.
Varying with respect to $\Lambda^{A\mu}$ ($e^J_{\mu}$ in \cite{P06}), however,
is quite subtle as one must respect $\Lambda^A_{\mu} \Lambda_{A\nu} = g_{\mu\nu}$.
That is, the variations $\delta \Lambda^{A\mu}$ and $\delta g_{\mu\nu}$ are not
independent. The tetrad field $e_{a\mu}$ thus implicitly also appears in \cite{P06}, but it
is fixed as a functional of the metric, see eq.\ (34) in \cite{P06}.
In this paper, however, it is crucial that $\delta e_{a \mu}$
and $\delta \Lambda^{Ca}$ are independent and that the full-fledged gauge
freedom of the tetrad is manifestly available.

The approach in this paper to a fully reduced canonical framework
is to eliminate all constraints and gauge degrees of freedom from the action.
After that, only the independent variables, which parametrize the constraint
surface, must be varied. The action is then transformed into canonical
form by certain variable transformations. A (3+1)-split with respect to a timelike
unit 4-vector $n_{\mu}$ with components $n_{\mu} = (-N, 0, 0, 0)$ or $n^{\mu} = (1, -N^i) / N$,
where $N$ is the lapse function and $N^i$ the shift vector,
most naturally fits to a fully reduced canonical formulation of gravity.
The three matter constraints can be solved as
\begin{align}
	np & \equiv n^{\mu} p_{\mu} = - \sqrt{m^2 + \gamma^{ij} p_{i} p_{j}} \,, \\
	nS_{i} & \equiv  n^{\mu} S_{\mu i}
		= \frac{p_{k} \gamma^{kj} S_{ji}}{np} = g_{ij} nS^{j} \,, \\
	\Lambda^{[j](0)} &= \Lambda^{[j](i)} \frac{p_{(i)}}{p^{(0)}} \,, \qquad
	\Lambda^{[0]a} = - \frac{p^a}{m} \,,
\end{align}
in terms of $p_i$, $S_{(i)(j)}$, and $\Lambda^{[i](k)}$, so one can put $\mathcal{L}_C=0$ now.
Here $\gamma^{ij}$ is the inverse of the induced 3-dim.\ metric $g_{ij} \equiv \gamma_{ij}$
of the hypersurfaces orthogonal to $n_{\mu}$.
A split of the Ricci rotation coefficients results in
\begin{align}
\omega_{kij} &= - \Gamma_{jik} + e_{i , k}^a e_{a j} \,, \\
n^{\mu} \omega_{k \mu i} &= K_{ki} - g_{ij} \frac{N^j_{,k}}{N}
	+ \frac{e_{ai}}{N} ( e^a_{0,k} - e^a_{l,k} N^l ) \,, \label{ric2} \\
\omega_{0ij} &= N K_{ij} - N_{j;i} + e_{i , 0}^a e_{a j} \,, \\
n^{\mu} \omega_{0 \mu i} &= K_{ij} N^j - N_{;i} - g_{ij} \frac{N^j_{,0}}{N}
	+ \frac{e_{ai}}{N} ( e^a_{0,0} - e^a_{l,0} N^l ) \,, \label{ric4}
\end{align}
where $_{;i}$ denotes the 3-dim.\ covariant derivative,
$\Gamma_{jik}$ the 3-dim.\ Christoffel symbols and
the extrinsic curvature $K_{ij}$ is given by
$2 N K_{ij} = - g_{ij,0} + 2 N_{(i;j)}$.
For convenience, we will immediately go to the time gauge $e^{\mu}_{(0)} = n^{\mu}$,
see \cite{S63}, also \cite{D62,K63,NT78}, as lapse and shift then
turn into Lagrange multipliers, like in the ADM formalism
(e.g., the $N^j_{,k} / N$ and $N^j_{,0} / N$ terms are
canceled in (\ref{ric2}) and (\ref{ric4})). It holds
\begin{align}
e_{i}^{(0)} &= 0 = e_{(i)}^0 \,, &
e^{(0)}_0 &= N = 1 / e_{(0)}^0 \,, \\
N^i &=  - N e^{i}_{(0)} \,, &
e^{(i)}_0 &= N^j e^{(i)}_j \,, \\
g_{ij} &= e_i^{(m)} e_{(m)j} \,, &
\gamma^{ij} &= e^i_{(m)} e^{(m)j} \,.
\end{align}
In passing we mention that in \cite{K72} a completely different tetrad
field has been chosen for a canonical formulation of test spinning particles
moving in gravitational fields.

The matter action in the covariant SSC $p^b S_{ab} = 0$ turns into
\begin{equation}
\mathcal{L}_M = \mathcal{L}_{MK} + \mathcal{L}_{MC} + \mathcal{L}_{GK}
	+ (\text{td}) \,,
\end{equation}
where $(\text{td})$ denotes an irrelevant total divergence.
The terms attributed to the kinetic matter part are given by
\begin{equation}
\begin{split}
\mathcal{L}_{MK} &= \bigg[ p_{i} + K_{ij} nS^j + A^{kl} e_{(j)k} e_{l,i}^{(j)} \nl
		- \bigg( \frac{1}{2} S_{kj}
			+ \frac{p_{(k} nS_{j)}}{np}
		\bigg) \Gamma^{kj}_{~~i} \bigg] \dot{z}^{i} \delta
	+ \frac{nS^i}{2 np} \dot{p}_i \delta \nl
+ \bigg[ S_{(i)(j)} + \frac{nS_{(i)} p_{(j)} -  nS_{(j)} p_{(i)}}{np} \bigg]
		\frac{\Lambda_{[k]}^{(i)} \dot{\Lambda}^{[k](j)}}{2} \delta \,,
\end{split}
\end{equation}
with $A^{ij}$ defined by
\begin{equation}
g_{ik} g_{jl} A^{kl} = \frac{1}{2} S_{ij} + \frac{nS_i p_j}{2 np} \,.
\end{equation}
The delta function $\delta$ is defined as $\int \mathrm{d}^3x \delta(x^i - z^i(t)) = 1$,
and the gauge $\tau=z^0=t$ was chosen, where $t$ is the time coordinate
of the object. A ``dot'' denotes the derivative with respect to $t$.
The complicated structure of these kinetic terms represents the
Dirac bracket arising from the covariant SSC.
The matter parts of the gravitational constraints result from
\begin{equation}
\mathcal{L}_{MC} = - N \mathcal{H}^{\text{matter}} + N^i \mathcal{H}^{\text{matter}}_i \,,
\end{equation}
with
\begin{align}
\mathcal{H}^{\text{matter}} &= - np \delta
	- K^{ij} \frac{p_i nS_j}{np} \delta - ( nS^k \delta )_{;k} \,, \\
\begin{split}
\mathcal{H}^{\text{matter}}_i &= (p_i + K_{ij} nS^j ) \delta \nl
	+ \bigg( \frac{1}{2} \gamma^{mk} S_{ik} \delta
		+ \delta_i^{(k} \gamma^{l)m} \frac{p_k nS_l}{np} \delta \bigg)_{;m} \,.
\end{split}
\end{align}
These coincide with the densitized projections $\mathcal{H}^{\rm matter} = \sqrt{\gamma}T_{\mu\nu} n^{\mu}n^{\nu}$
and $\mathcal{H}^{\rm matter}_i = - \sqrt{\gamma}T_{i \nu} n^{\nu}$
of the stress-energy tensor in covariant SSC, eq.\ (\ref{SET}), see also
\cite{SSH08}, where $\gamma = \text{det}(\gamma_{ij})$.
Further, some terms attributed to the kinetic part of the gravitational
field appear as
\begin{equation}
\mathcal{L}_{GK} = A^{ij} e_{(k)i} e_{j,0}^{(k)} \delta \,.
\end{equation}

Now we proceed to Newton-Wigner (NW) variables $\hat{z}^i$, $P_i$, $\hat{S}_{(i)(j)}$, and $\hat{\Lambda}^{[i](j)}$,
which turn the kinetic matter part $\mathcal{L}_{MK}$ into canonical form.
The variable transformations read
\begin{align}
z^i &= \hat{z}^i - \frac{nS^i}{m - np} \,, \quad nS_i = - \frac{p_k \gamma^{kj} \hat{S}_{ji}}{m} \,, \\
S_{ij} &= \hat{S}_{ij} - \frac{p_i nS_{j}}{m - np} + \frac{p_j nS_i}{m - np} \,, \\
\Lambda^{[i](j)} &= \hat{\Lambda}^{[i](k)} \bigg( \delta_{kj} + \frac{p_{(k)}p^{(j)}}{m (m - np)} \bigg) \,, \\
\begin{split}
P_i &= p_i + K_{ij} nS^j + \hat{A}^{kl} e_{(j)k} e_{l,i}^{(j)} \nl
	- \bigg( \frac{1}{2} S_{kj}
		+ \frac{p_{(k} nS_{j)}}{np}
	\bigg) \Gamma^{kj}_{~~i} \,, \label{NWmom}
\end{split}
\end{align}
where $\hat{A}^{ij}$ is given by
\begin{equation}
g_{ik} g_{jl} \hat{A}^{kl} = \frac{1}{2} \hat{S}_{ij} + \frac{m p_{(i} nS_{j)}}{np (m-np)} \,.
\end{equation}
The NW variables have the important properties $\hat{S}_{(i)(j)} \hat{S}_{(i)(j)} = 2 s^2 = \text{const}$
and $\hat{\Lambda}_{[k]}^{(i)} \hat{\Lambda}^{[k](j)} = \delta_{ij}$,
which implies that
$\mathrm{d} \hat{\theta}^{(i)(j)} \equiv \hat{\Lambda}_{[k]}^{(i)} \mathrm{d} \hat{\Lambda}^{[k](j)}$
is antisymmetric. The redefinitions of position, spin tensor, and angle-type
variables are actually quite natural generalizations of the
Minkowski space versions, cf., refs.\ \cite{HR74,F65}, to curved spacetime.
However, there is no difference between linear momentum $p_i$ and canonical momentum
$P_i$ in the Minkowski case.
In these NW variables, one has
\begin{equation}
\mathcal{L}_{GK} + \mathcal{L}_{MK} = \hat{\mathcal{L}}_{GK} + \hat{\mathcal{L}}_{MK}
	+ (\text{td}) \,,
\end{equation}
with (from now on $\delta = \delta(x^i - \hat{z}^i(t))$)
\begin{align}
\hat{\mathcal{L}}_{MK} &= P_i \dot{\hat{z}}^i \delta + \frac{1}{2} \hat{S}_{(i)(j)} \dot{\hat{\theta}}^{(i)(j)} \delta \,, \\
\hat{\mathcal{L}}_{GK} &= \hat{A}^{ij} e_{(k)i} e_{j,0}^{(k)} \delta \,.
\end{align}
Notice that all $\dot{p}_i$ terms in the action have been canceled
by the redefinition of the position. Further, all $K_{ij}$ terms were
eliminated from $\mathcal{L}_{MC}$ and $\mathcal{L}_{MK}$ by
the redefinition of the linear momentum.
If the terms explicitly depending on the triad $e^{(i)}_j$ are
neglected, the known source terms of Hamilton and momentum
constraints in canonical variables, respectively (4.23) and
(4.25) in \cite{SSH08}, are obtained.

The final step goes with the ADM action functional of the gravitational
field, \cite{ADM62,DW67,RT74}, but in tetrad form
as derived in \cite{DI76}. In passing we mention
that Kibble, \cite{K63}, was applying the Schwinger canonical formalism
which uses a different set of field variables, \cite{S63}.
The canonical momentum conjugate to $e_{(k)j}$ is given by
\begin{equation}
\bar{\pi}^{(k)j} = 8\pi \frac{\partial \mathcal{L}}{\partial e_{(k)j,0}}
= e_i^{(k)} \pi^{ij} + e_i^{(k)} 8\pi \hat{A}^{ij} \delta \,,
\end{equation}
where
\begin{equation}
\pi^{ij} = \sqrt{\gamma} (\gamma^{ij}\gamma^{kl} - \gamma^{ik}\gamma^{jl})K_{kl} \,.
\end{equation}
Legendre transformation leads to
\begin{equation}
\hat{\mathcal{L}}_{GK} + \mathcal{L}_G
	= \frac{1}{8\pi} \bar{\pi}^{(k)j} e_{(k)j,0}
        - \frac{1}{16\pi} \mathcal{E}_{i,i} + \mathcal{L}_{GC} + (\text{td}) \,.
\end{equation}
The explicit form of the non-irrelevant total divergence
$\mathcal{E}_{i,i}$ emerges from the total divergence that can be added to the action, see eq.\ (\ref{EH}).
As shown in \cite{ADM62,DW67,RT74} by different methods, see also \cite{S63},
it must be given by $\mathcal{E}_i = g_{ij,j} - g_{jj,i}$ for asymptotically flat spacetimes.
The total energy reads $E = \frac{1}{16\pi}\oint d^2 s_i \mathcal{E}_i$.
The constraint part takes the form
\begin{equation}
\mathcal{L}_{GC} = - N \mathcal{H}^{\text{field}} + N^i \mathcal{H}^{\text{field}}_i \,,
\end{equation}
with
\begin{gather}
	\mathcal{H}^{\text{field}} = - \frac{1}{16\pi\sqrt{\gamma}} \left[ \gamma \text{R}
		+ \frac{1}{2} \left( g_{ij} \pi^{ij} \right)^2
		- g_{ij} g_{k l} \pi^{ik} \pi^{jl}\right] , \\
	\mathcal{H}^{\text{field}}_i = \frac{1}{8\pi} g_{ij} \pi^{jk}_{~~ ; k} \,,
\end{gather}
where $\text{R}$ is the 3-dim.\ Ricci scalar.
Due to the symmetry of $\pi^{ij}$, not all components of $\bar{\pi}^{(k)j}$ are independent variables
(i.e., the Legendre map is not invertible), leading to the additional constraint
$\bar{\pi}^{[ij]} = 8\pi \hat{A}^{[ij]} \delta$.
This constraint will be eliminated by going to the spatial symmetric gauge
$e_{(i)j} = e_{ij} = e_{ji}$ , $e^{(i)j} = e^{ij} = e^{ji}$.
Then the triad is fixed as the matrix square-root of the 3-dim.\ metric, $e_{ij} e_{jk} = g_{ik}$, or
\begin{equation}
e_{ij} = \sqrt{(g_{kl})} \,.
\end{equation}
Therefore, we can define a quantity $B^{kl}_{ij}$ as
\begin{equation}
e_{k[i} e_{j]k,\mu} = B^{kl}_{ij} g_{kl,\mu} \,,
\end{equation}
or, in explicit form,
\begin{equation}
2 B^{kl}_{ij} = e_{mi} \frac{\partial e_{mj}}{\partial g_{kl}} - e_{mj} \frac{\partial e_{mi}}{\partial g_{kl}} \,.
\end{equation}
This expression may be evaluated perturbatively, cf.\ \cite{SSH08}.
It holds $B^{kl}_{ij} \delta_{kl} = 0$. Furthermore,
\begin{equation}
e_{(k)i} e_{j,\mu}^{(k)} = B^{kl}_{ij} g_{kl,\mu} + \frac{1}{2}
g_{ij,\mu}\,,  \label{eliminatee}
\end{equation}
which gets applied as
\begin{equation}
\bar{\pi}^{(k)j} e_{(k)j,0} = \frac{1}{2} \pi_{\text{can}}^{ij} g_{ij,0} \,,
\end{equation}
with the new canonical field momentum
\begin{equation}
\pi_{\text{can}}^{ij} = \pi^{ij} + 8\pi \hat{A}^{(ij)} \delta + 16\pi B^{ij}_{kl} \hat{A}^{[kl]} \delta \,. \label{pican}
\end{equation}
Notice that (\ref{eliminatee}) can also be used to replace the triad terms in (\ref{NWmom}).

Next the gravitational constraints arising from the variations $\delta N$ and $\delta N^i$,
\begin{equation}
\mathcal{H}^{\text{field}} + \mathcal{H}^{\text{matter}} = 0 \,, \quad
\mathcal{H}^{\text{field}}_i + \mathcal{H}^{\text{matter}}_i = 0 \,, \label{Gconst}
\end{equation}
are eliminated by also imposing the gauge conditions
\begin{align}
3 g_{ij,j} - g_{jj,i} = 0 \,, \quad
\pi^{ii}_{\text{can}} = 0 \,,
\end{align}
which allow for the decompositions
\begin{align}
	g_{ij} = \Psi^4 \delta_{ij} + h^{\text{TT}}_{ij} \,, \quad
	\pi^{ij}_{\text{can}} = \tilde{\pi}^{ij}_{\text{can}} + \pi^{ij\text{TT}}_{\text{can}} \,, \label{ADMTT}
\end{align}
where $h^{\text{TT}}_{ij}$ and $\pi^{ij\text{TT}}_{\text{can}}$ are transverse traceless,
e.g., $h^{\text{TT}}_{ii} = h^{\text{TT}}_{ij,j}=0$, and $\tilde{\pi}^{ij}_{\text{can}}$ is
related to a vector potential $V^i_{\text{can}}$ by
$\tilde{\pi}^{ij}_{\text{can}} = V^i_{\text{can} , j} + V^j_{\text{can} , i} - \frac{2}{3} \delta_{ij} V^k_{\text{can} , k}$.
The gravitational constraints can now be solved for
$\Psi$ and $\tilde{\pi}^{ij}_{\text{can}}$, leaving $h^{\text{TT}}_{ij}$
and $\pi^{ij\text{TT}}_{\text{can}}$ as the final degrees of freedom
of the gravitational field.
Notice that our gauge condition $\pi^{ii}_{\text{can}} = 0$ deviates
from the original ADM one $\pi^{ii} = 0$ by spin corrections at
5PN. The action reads, with $16\pi$-abuse of canonicity,
\begin{equation}
W = \int \mathrm{d}^4 x \frac{\pi^{ij\text{TT}}_{\text{can}}}{16\pi} h^{\text{TT}}_{ij,0}
	+ \int \mathrm{d}t \bigg[ P_i \dot{\hat{z}}^i + \frac{1}{2} \hat{S}_{(i)(j)} \dot{\hat{\theta}}^{(i)(j)} - E \bigg] ,
\end{equation}
and is in fully reduced canonical form. The dynamics is completely
described by the ADM energy $E$, which turns into the volume integral
\begin{equation}
E = - \frac{1}{2\pi} \int \mathrm{d}^3 x \Delta \Psi[\hat{z}^i, P_i, \hat{S}_{(i)(j)}, h^{\text{TT}}_{ij}, \pi^{ij\text{TT}}_{\text{can}}] \,,
\end{equation}
and is the total Hamiltonian ($E=H$), once it is expressed in terms of the canonical
variables by solving (\ref{Gconst}) with (\ref{ADMTT}).

The equal-time Poisson bracket relations take the standard form (${\bf x}= (x^i))$,
\begin{gather}
\{ \hat{z}^i, P_j\} = \delta_{ij} \,, \quad \{{\hat S}_{(i)},
{\hat S}_{(j)}\} =
\epsilon_{ijk} {\hat S}_{(k)} \,, \quad \\
\{h^{\text{TT}}_{ij}({\bf x},t), \pi^{kl\text{TT}}_{\text{can}}({\bf x}',t)\}
= 16\pi \delta^{\text{TT}kl}_{ij}\delta({\bf x} - {\bf x}') \,,
\end{gather}
zero otherwise, where
${\hat S}_{(i)}= \frac{1}{2}\epsilon_{ijk}{\hat S}_{(j)(k)}$,
$\epsilon_{ijk}=(i-j)(j-k)(k-i)/2$,
and $\delta^{\text{TT}ij}_{mn}$ is the TT-projection operator, see, e.g., \cite{SSH08}.
The Hamiltonian $H$ generates the time evolution in the reduced
matter-field phase space.
Generalization and application to many-body systems is quite straightforward, see \cite{SSH08}.
The total linear ($P_i^{\text{tot}}$) and angular ($J_{ij}^{\text{tot}}$) momenta take the
forms (particle labels are denoted $a$),
\begin{align}
	P_i^{\text{tot}} &= \sum_a P_{ai}
		- \frac{1}{16\pi} \int \mathrm{d}^3x \, \pi_{\text{can}}^{kl\text{TT}} h^{\text{TT}}_{kl,i} \,, \label{Ptot} \\
\begin{split}
	J_{ij}^{\text{tot}} &= \sum_a ( \hat{z}_a^i P_{aj} - \hat{z}_a^j P_{ai}  + \hat{S}_{a(i)(j)}) \nl
        - \frac{1}{8\pi} \int \mathrm{d}^3x \, ( \pi_{\text{can}}^{ik\text{TT}} h^{\text{TT}}_{kj}
			- \pi_{\text{can}}^{jk\text{TT}} h^{\text{TT}}_{ki} ) \nl
		- \frac{1}{16\pi} \int \mathrm{d}^3x \, ( x^i \pi_{\text{can}}^{kl\text{TT}} h^{\text{TT}}_{kl,j} - x^j \pi_{\text{can}}^{kl\text{TT}} h^{\text{TT}}_{kl,i} ) \label{Jtot} \,,
\end{split}
\end{align}
and are obtained from the reduced action in the standard Noether manner.

Correctness of the developed formalism has been explicitly checked through
conservative 3PN and dissipative 3.5PN orders (or conservative 3.5PN and
dissipative 4PN orders for spin-orbit interaction if spin is counted of order $1/c$) by an
independent method based on the full Einstein field equations with
the Tulczyjew-Dixon stress-energy tensor as source term, \cite{SSH08,SW09}.
Further, an alternative derivation of $P_i^{\text{tot}}$ and $J_{ij}^{\text{tot}}$
via surface integrals and the momentum constraint is given in \cite{SW09}, which
provides a check of the canonicity of the variables up to all orders.

\textbf{Note added in v3}:
Extensions of the action in \cite{HR74} to gravitational interactions have already been
considered in \cite{BI75}. Whereas we have only given a minimal coupling here,
Eq.\ (\ref{spin1}), in \cite{BI75} even nonminimal couplings to gravity
and couplings to the electromagnetic field were discussed.
However, a separation into independent variations for field
$\delta e_{a \mu}$ and matter $\delta \Lambda^{Ca}$ as necessary
in the present paper was not considered.

\acknowledgments
We thank S.\ Hergt for helpful discussions.
This work is supported by the Deutsche Forschungsgemeinschaft (DFG) through
SFB/TR7 ``Gravitational Wave Astronomy'' and GRK 1523.

\end{document}